\documentclass[article,10pt]{elsarticle}

 \usepackage{amssymb}
 \usepackage{txfonts,color}
\usepackage{graphicx,multicol}
\journal{Astroparticle Physics}

\begin{document}

\begin{frontmatter}

\title{On the Detectability of High-Energy\\ Galactic Neutrino Sources}

\author{Francesco Vissani}
\address{Laboratori Nazionali del Gran Sasso, Assergi (AQ), Italy}
\author{Felix Aharonian} 
\address{Dublin Institute for Advanced Studies, 
31 Fitzwilliam Place, Dublin 2, Ireland}
\author{Narek Sahakyan}
\address{Universit\`a di Roma ``La Sapienza''  and ICRANet, Pescara, Italy}
\date{}     
\small
\begin{abstract}
With the arrival of km$^3$ volume scale neutrino detectors
the chances to detect the first astronomical sources
of TeV neutrinos will be dramatically increased. 
While the theoretical estimates of the neutrino  fluxes contain large uncertainties, 
we  can formulate the conditions for the 
detectability of certain neutrino sources phenomenologically.
In fact, since most galactic neutrino sources 
are transparent for TeV $\gamma$-rays, 
their detectability implies 
a minimum flux of the accompanying $\gamma$-rays.
For a typical energy-dependence of detection areas  of
km$^3$ volume neutrino detectors, we obtain the 
quantitative condition $I_{\gamma}(20\mbox{ TeV})>2 \times 10^{-15} \ \rm ph/cm^2 s$, 
that thanks to the normalization of the $\gamma$-ray spectrum at 20 TeV
appears to be quite robust, i.e.\ almost independent of the shape of energy spectrum of
neutrinos.
We remark that this condition is satisfied by the young
supernova remnants RX J1713.7-3946 and  RX J0852.0-4622 (Vela Jr) - two of the strongest
galactic $\gamma$-ray sources.
The preliminary condition for the detectability of high energy neutrinos
is that the bulk of $\gamma$-rays has a hadronic origin:
A new way to test this hypothesis for  RX J1713.7-3946 is proposed. 
Finally, we assess the relevance of a neutrino
detector located in the Northern Hemisphere for the search for
galactic neutrino sources. In particular, we argue that if the
TeV neutrino sources correlate with the galactic mass distribution,
the probability that some of them will be observed by  
a detector in the Mediterranean Sea is larger 
by a factor of 1.4-2.9 compared to the one  of IceCube.

%
\end{abstract}
               \end{frontmatter}

\begin{twocolumn}

\section{Introduction}
The search for neutrinos with $E_\nu>\mbox{TeV}$ (Lipari 2006 \cite{lip}) with telescopes of volumes 
in the km$^3$ scale is  considered important. IceCube is collecting  exposures of the order of km$^2\times$year and this will be continued and extended by KM3NeT.\footnote{All considerations below apply to any large (i.e., with volumes 
of the order of one km$^3$)
neutrino telescope 
of the Northern hemisphere; KM3NeT is taken as an example, being the most 
advanced project of this type at present.}  
As has happened in the past ({e.g.}, for X-ray searches) the new instruments 
could eventually  lead to surprising outcomes.  
The hope for surprises is certainly one strong motivation of the search 
for the sources of high energy neutrinos that plausibly are also sources of  cosmic rays.
At the same time, there are many reasons why we would like 
to have defined expectations on high energy neutrinos: 
to interpret the results of the observations, 
to plan the future research, to better focus our goals, 
to optimize the new instruments.
The trouble is that the hypotheses 
on which the present expectations are based 
are still rather uncertain and difficult to test.
Thus we do not have reliable predictions yet, 
and this limits our capability to plan the next steps.

In this paper, we focus on this aspect of the 
search for high energy neutrino sources that we are now beginning. 
We discuss several aspects:
We emphasize the relevance of $\gamma$-ray
observations in the 10-100 TeV energy range for high energy neutrinos; we 
analyze the prospect to understand better certain $\gamma$-ray sources 
that have a special theoretical interest in connection with high energy neutrinos; 
we clarify the argument in favor of a neutrino telescope in the Northern hemisphere.
We focus on the subclass of galactic sources that are 
of particular interest for future instruments located in the Northern hemisphere.

The outline of this paper is as follows. 
First, considering the high energy $\gamma$-ray 
sources that are transparent to the radiation, 
we characterize those of them  that could be, at the same time,  
bright enough neutrino sources (Sec.~\ref{s2}).
Next, after recalling the relevant theoretical context, 
we discuss the prospects of obtaining more defined 
expectations for one of the most interesting of these $\gamma$-ray 
sources, namely, 
the young supernova remnant named RX J1713.7-3946 (Sec.~\ref{s3}).
Finally,  we quantify in Sec.~\ref{s1}  
the importance of monitoring the Southern high energy neutrino sky on the
basis of the Galactic matter distribution.

\section{Using high energy
$\gamma$-rays as a guide for high energy neutrino search
\label{s2}}
Here we consider a precise assumption on the astrophysical neutrino 
sources: We suppose that they are transparent to the very high energy gamma radiation.
In this way we can derive upper limits on  neutrinos, by postulating that {\em all} 
$\gamma$-rays originate from proton-proton collisions. 
In fact, the yield of neutral mesons and of charged mesons are strictly connected
and there is a linear relation between the fluxes of high 
energy $\gamma$-rays and neutrinos (Vissani 2006 \cite{Vissa}, Villante \& Vissani 2008~\cite{vv}).

\bigskip

\begin{table}[t]
\centerline{
\begin{tabular}{r|ccccccc}
$N_\gamma\ \ \ \ \ \ \ \ \ \ \ $&&&&$E_c=$ &&&\\[1ex]
& {$10^0$} & {$10^{0.5}$} & {$10^1$} & {$10^{1.5}$} & {$10^2$} & {$10^{2.5}$} & {$10^3$}  \\
  \hline 
{1.8} &  70 &  16  & 4.9 & 2.1 & 1.1 & 0.7 & 0.5 \\
{1.9} &  86 &  20  & 6.7 & 3.0 & 1.7 & 1.1 & 0.8 \\
$\alpha=\ \ \ \ $ {2.0} &  110 & 25 & 9.0 & 4.2 & 2.5 & 1.7 & 1.3 \\
{2.1} &  130 & 32 & 12  & 5.9 & 3.5 & 2.5 & 2.0 \\
{2.2} &  160 & 41 & 16  & 8.0 & 5.0 & 3.6 & 3.0 
\end{tabular}}
\caption{\em
Normalization of the $\gamma$-ray fluxes {$N_\gamma$},  in units of 
$10^{-12} / {\mbox{\rm (cm$^2$ s TeV)}} $ that corresponds to an induced flux $I_{\mu+\bar\mu}(>1\mbox{\rm TeV})=1/(\mbox{\rm km}^2\mbox{\rm yr})$.
First column: slope of the $\gamma$-ray flux, {$\alpha$}. 
First row: cutoff energy of the $\gamma$-ray spectrum, {$E_{c}$}, measured in TeV.
See Eq.~\ref{de}.
\label{tab1}}
\end{table}

We can then quantify the concept of ``promising'' $\gamma$-ray sources. 
Suppose that the $\gamma$-ray flux has the form:
\begin{equation}
I_\gamma(E_\gamma)={N_\gamma}\times (E_\gamma/\mbox{1 TeV})^{-{\alpha}}\times \exp\left[-\sqrt{E_\gamma/{E_{c}}}\right],
\label{de}
\end{equation}
where we consider the ranges of parameters: 
$\alpha= 1.8 - 2.2$ (=slope) and $E_c = 1$ TeV$-$1 PeV (=energy cutoff). 
This form corresponds to an exponential cutoff in the 
spectrum of the cosmic rays that generate the $\gamma$-rays, see 
Kappes {et al.}\  2007 \cite{12}, and has been tested for 
adequacy on the available $\gamma$-ray data. 
Following Villante \& Vissani 2008 \cite{vv}, and 
requiring an induced flux of {1 muon} or antimuon
per km$^2$ per year above 1 TeV ({i.e.}\ 1 signal event in a 
conventional neutrino telescope with an exposure of 1 km$^2\times$yr),
we determine $N_\gamma$ for each value of $\alpha$ and $E_c$; the results
are given in Tab.~\ref{tab1} and are further illustrated in Fig.~\ref{fig2}.

This table and this figure 
identify the transparent $\gamma$-ray sources that could be 
interesting neutrino sources. There is a wide variety of possibilities, 
ranging from intense $\gamma$-ray sources to weak ones; this is due to the 
fact that the main contribution to the neutrino signal is from energies larger than 
1 TeV, where we still have limited information from $\gamma$-ray observations. 

By a systematic exploration of 
the $\gamma$-ray sky till  $E_\gamma\sim 100$ TeV  and of the sources with an intensity 
above $10^{-12} /$ (cm$^2$ s TeV) at 1 TeV, 
we could have a guide for the search of very high energy neutrinos, at least for the sources that are transparent to their $\gamma$-ray radiation.
It is interesting to note that at 20 TeV, all fluxes of Tab.~\ref{tab1} are in the narrow range 
\begin{equation}
I_\gamma(20\ \rm TeV)=(2-6) \times 10^{-15}\ \mbox{ph}/\ ({\rm cm^{2}\ s\ TeV})
\label{cat}
\end{equation}
that characterizes the region of energies and of intensities 
where the $\gamma$-ray observations are more relevant for the high energy 
neutrino detectors: see again Fig.~\ref{fig2}.

We remark that Eq.~\ref{cat} is a new result.
Its relevance can be understood better by recalling 
that the existing $\gamma$-ray detectors 
have explored mostly the region of energy around 
the TeV. Thus, the future $\gamma$-ray measurements in the region 10-100 TeV--e.g.\ 
those by the Cherenkov Telescope Array (CTA) instrument--will 
have an important impact on the expectations of high energy 
neutrinos.
In summary, it will be possible to clarify the expectations of neutrino astronomy,
by the measurements of future $\gamma$-ray observatories.

\begin{figure}[t]
\centerline{\includegraphics[width=0.5\textwidth,angle=0]{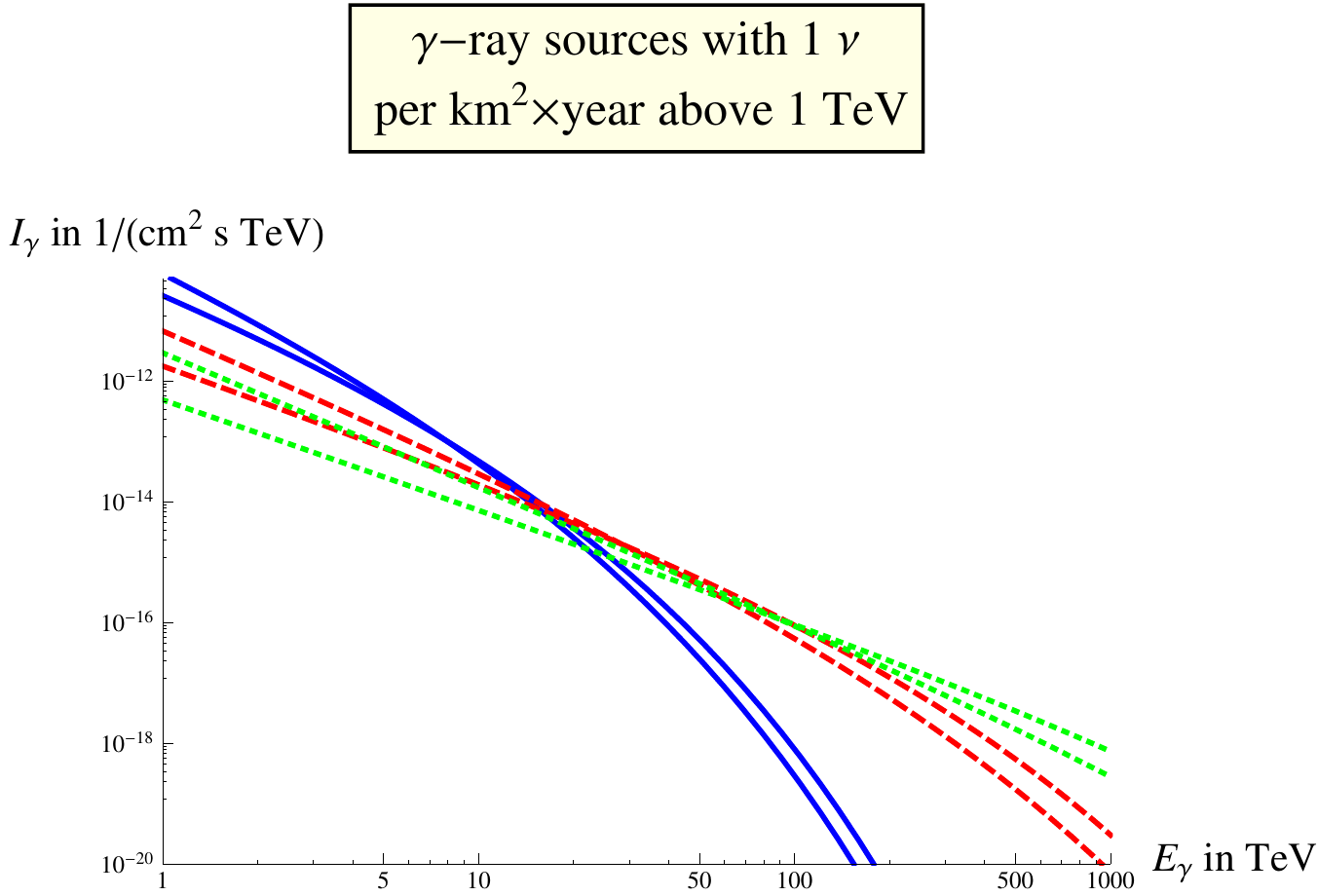}}
\caption{\em Fluxes of $\gamma$-rays corresponding to 1 event above 1 TeV 
in a neutrino telescope with an exposure of 1 km$^2\times$yr, as given 
in Eq.~\ref{de} and Tab.~\ref{tab1}. We selected the values 
$\alpha=1.8$ and $2.2$ (the first being smaller at low energies and larger at high energies) 
and $E_c=1,10^{1.5}$ and $10^3$ TeV (continuous, dashed and dotted lines, respectively).  
\label{fig2}}
\end{figure}

\paragraph{SNR as a major example of transparent $\gamma$-ray source}
Shell type supernova remnants (SNR) are an important example of 
astronomical sources of $\gamma$-rays that is  
expected to be transparent to their $\gamma$-ray emission.
A few young SNR, recently observed in the 
TeV range,\footnote{A useful resource is 
the HESS Source Catalog that can be consulted at: 
{\tt http://www.mpi-hd.mpg.de/hfm/HESS/pages/home/sources/}.}  
are known to exceed the bound in Eq.~\ref{cat}, 
thus being of particular interest.
We discuss them here to illustrate the issue further: 
\begin{itemize}
\item
The first  example is the supernova remnant called  RX J1713.7-3946 and measured by
HESS, Aharonian {et al.}\ 2007 \cite{hess}. 
It has a $\gamma$-ray spectrum 
that is reasonably well described assuming $\alpha=1.79\pm 0.06$ and $E_c=3.7\pm 1$ TeV in Eq.~\ref{de}
(Villante \& Vissani 2007 \cite{vvvv}) and that 
has an 
intensity at 20 TeV of $(1.7\pm 0.3)\times 10^{-14}\ /\ ({\rm cm^{2}\ s\ TeV})$.
Correspondingly,  the maximum value of the neutrino signal from this source is 
larger than 1 event per km$^2$ per year, and more precisely we have:
\begin{equation}
I_{\mu+\bar{\mu}}(>1\mbox{ TeV}) = ( 2.4\pm 0.3 \pm 0.5 )\  /\ ({\rm km^{2}\ yr})
\label{cacat}
\end{equation}
see Tab.~2 and Sect.~IV of Villante \& Vissani 2008~\cite{vv}. 
\item
A second example is the SNR called 
Vela Jr  (RX J0852. 0-4622) as observed in Aharonian {et al.}\ 2007b \cite{hessb}.
The available $\gamma$-ray observations of this object are less complete.
Its measured spectrum has been described simply by a 
power law and its emission at 20 TeV is in the range 
$(1-3)\times 10^{-14}\ /\ ({\rm cm^{2}\ s\ TeV})$; however, 20 TeV is the 
highest measured energy.
\end{itemize}
Note that both these SNR's 
are in the South $\gamma$-ray sky, and are thus potentially
interesting for neutrino telescopes located in the Northern hemisphere (see Sec.~\ref{s1}). 
In the next section,  we recall the reasons why such SNR's 
are considered interesting and discuss in more details 
the present understanding of RX J1713.7-3946 and 
the prospects of improvement.

\paragraph{Remarks and caveats}
We did not include latitude dependent effects, such as the limited time to observe a source 
(discussed in the last section) or the absorption in the Earth, in order to simplify the discussion.
The latter effect is more severe for the fluxes that 
extend up to the highest energies, namely those with a smaller 
value of $\alpha$ and/or a higher energy cutoff. We roughly take into account this, 
by limiting the spectrum to $E_\nu<1$~PeV.

Let us repeat that the $\gamma$-ray data permit us to obtain an upper bound 
on neutrinos, postulating that the source is transparent to its $\gamma$-rays. But
in some cases,  high energy neutrinos and $\gamma$-rays do not correlate. 
This is thought to happen for certain interesting astrophysical objects such as 
galactic binary systems containing a luminous optical star and a compact object
(microquasars) where the absorption of $\gamma$-rays is considerable.
Cases like this increase the {\em a priori} chance 
of having surprisingly large neutrino fluxes.
At the same time, such a situation
causes further difficulties to obtain reliable 
expectations, due to the increased dependence on an 
uncertain theoretical modelling.
For a more complete discussion of alternative galactic 
neutrino sources, see Aharonian 2007~\cite{aaaaa}.

\section{Toward reliable predictions for the SNR RX J1713.7-3946
\label{s3}}
In this section we analyze why supernova remnants are expected to act 
as emitters of high energy neutrinos and discuss in details the promising case of the supernova remnant 
called  RX J1713.7-3946.
\bigskip

The kinetic energy of supernova remnants  (SNR) is one order of magnitude larger than
the cosmic ray losses of the Galaxy (Ginzburg \& Syrovatskii 1964 \cite{gs}) 
and diffusive acceleration on the shock wave
(Fermi 1949 \cite{fermi}) can provide the mechanism for cosmic-ray
acceleration (see {e.g.}, Malkov \& O'Drury 2001 \cite{teo}):
Thus, we expect that SNR's contain high densities of cosmic rays.
The SNR can also be sources of high energy $\gamma$-rays and neutrinos, especially when they are  
associated with molecular clouds, that act as a target for the cosmic ray collisions  
(Aharonian, O'Drury, V\"olk 1994 \cite{aha1}, O'Drury, Aharonian, V\"olk 1994~\cite{aha2}). 

The spectrum of the young SNR RX J1713.7-3946, 
measured by HESS (Aharonian {et al.}\ 2007 \cite{hess}) till 100 TeV and (as already discussed) 
exceeding the bound of Eq.~\ref{cat}, has a special interest in
the discussion of high energy neutrinos.
This is even more true when one realizes that this SNR interacts 
with a system of molecular clouds detected by NANTEN (Fukui {et al.}\ 2003 \cite{nanten}, 
Sano {et al.}\ 2010 \cite{sano}). It is in the Southern neutrino sky,
relatively close to us, $D\sim 1$~kpc.

These are the reasons why it is urgent to ask: 
How far we are from understanding the high energy neutrinos of {\em this} SNR? To address 
this question, we have to consolidate the physical picture of RX J1713.7-3946, that 
can be done only employing in the best way the available (theoretical and observational) 
information.  

\paragraph{A model for the spectrum}

A related question that we should tackle 
is which is the nature of the electromagnetic spectrum of
RX J1713.7-3946. 
There are various models  in the literature, {e.g.}, 
Malkov {et al.}\ 2005 \cite{mea}, 
Berezhko \& V\"olk 2006-2010 \cite{bv},
Morlino {et al.}\ 2009 \cite{blabla}, Zirakashvili \& Aharonian 2010 \cite{az},
Ellison {et al.}\ 2010 \cite{s}, Fan {et al.}\ 2010 \cite{fan}. In a typical model, 
the $\gamma$-ray emission is dominated by a single 
mechanism at all energies, which reduces the question of 
neutrinos to a dichotomy.

We focus on one proposal of Zirakashvili \& Aharonian 2010 \cite{az},
where the spectrum is instead composite: it has significant contributions both from the Inverse Compton (IC, {i.e.}\ leptonic mechanism) and from neutral pion decays ($\pi^0$, {i.e.}\ hadronic mechanism). 
Even if their model will turn out to be incorrect in some quantitative aspect,  such a hypothesis 
allows us to make one step ahead in the right direction: to understand neutrinos, 
we need to know {\em which part} of the $\gamma$-ray emission is hadronic. 

This proposal has good physical motivations:
1)~The similarity of the features observed in X-rays is explained, since  
the IC dominates the integrated 
spectrum measured by HESS. 
2)~Attributing the high energy tail of the spectrum to $\pi^0$ decays instead
overcomes the difficulties in accounting for it by IC, whose spectrum 
should be cut-off abruptly.
The question we want to address becomes: 
How do we test the predictions of this model?

\paragraph{Prospects of observational tests}

The measurements of Agile and Fermi in the energy range 1-10 GeV
will be a key test (see {e.g.}, Morlino, Blasi, Amato 2009 \cite{blabla}),
for the shape of the $\gamma$-ray spectrum at GeV energies 
depends on the mechanism of emission:
Assuming that protons and electrons have power-law 
spectra with the same slope, $\propto E^{-\alpha}_{p,e}$,  
$\pi^0$ decays give $\propto E_\gamma^{-(\alpha-0.1)}$ whereas IC gives 
$\propto E_\gamma^{-(\alpha+1)/2}$; thus,
extrapolating the $\gamma$-ray spectra
from the lowest point measured by HESS, $E_\gamma=300$ GeV, 
we find that the hadronic mechanism leads to an emission 3 times more intense 
than the one due to the leptonic mechanism
already at 10 GeV. But it is unclear whether Agile or Fermi 
will attain sufficient angular resolution to reveal that the $\gamma$-rays come preferentially
from the sites where cosmic ray collisions
and $\pi^0$ decays are more frequent, {i.e.}\ the molecular clouds.

This qualifying hypothesis could be verifiable 
at much larger energies.  In fact, the model of Zirakashvili \& Aharonian 2010 \cite{az}
predicts, at several tens of TeV,  a $\gamma$-ray signal enhanced in the direction of the overdense
molecular clouds of NANTEN, of size $(2-8)\ \mu$sr  (Sano {et al.}\ 2010 \cite{sano}), 
on top of the known background distribution due to misidentified cosmic rays and of a minor 
component of the signal distributed as the SNR shell.

\begin{table}[t]
\centerline{\small
\begin{tabular}{c|c|c|c}
Energy  & Dominant & Observational & Relevant \\[0ex]
of $\gamma$-rays  & emission & test   & data\\ \hline 
$\sim\!1\!-\!10$ GeV  & $\pi^0$  & intensity \& shape & Fermi, Agile \\ 
$\sim\!1\!-\!10$ TeV  & IC  & SNR shell & HESS\\
$>10$ TeV  & $\pi^0$ & molecular clouds & HESS  \\
\end{tabular}}
\caption{\em Tests of the Zirakashvili \& Aharonian model for the $\gamma$-ray 
spectrum of the SNR named RX J1713.7-3946. First column, the energy range 
of the measurement; second, the dominant mechanism of emission expected in the model;
third, the possible test; last column, the relevant experiment. See the text for details. 
\label{ttb2}}
\end{table}

Do we have enough data to test this picture? 
HESS (Aharonian {et al.}\ 2007 \cite{hess}) has 1021 (resp., 474) 
events ON against 751 (resp., 338) OFF above 20 (resp., 30) TeV, namely 
about 250 (resp., 130)  signal events\footnote{The terminology ON/OFF refers to the two cases 
when the gamma ray telescope points to the source and when instead it 
points to a region where no signal is expected.}.
To illustrate the meaning of these numbers, suppose that 
750 background events are uniformly distributed in 25 patches of equal area; 
thus, 30 signal events in a single patch are enough to double the 
average density of events.
The low statistics conditions suggest
an unbinned likelihood analysis of the $\gamma$-ray data, as those 
proposed  for similar applications in neutrino astronomy 
by Braun {et al.}\ 2008 \cite{monta} and 
by Ianni {et~al.}~2009~\cite{ianni}.

A r\'esum\'e of possible tests is provided in Tab~\ref{ttb2}.

The above estimates show that the existing HESS data can only marginally
provide a decisive  study of energy dependent $\gamma$-ray morphology 
of RX J1713.7-3946. Thus, it is highly desirable to increase significantly 
the TeV photon statistics by new observations of the source. 
Presently such observations can be performed only by the HESS
array of telescopes. However, because of the limited potential of HESS at
energies above 10 TeV,  we can hope for enhancement of  photon statistics,
for any reasonable observation time, only by a factor of two or so. A real
breakthrough in this regard is  expected  only with the next generation
$\gamma$-ray instruments like CTA.


\paragraph{Implications for high energy neutrino astronomy}
If the model of Zirakashvili \& Aharonian
will be validated by future data analyses, the induced muon flux 
from RX J1713.7-3946 will be
lower than the upper limit  that we derive in the extreme hypothesis of hadronic emission
from Eq.~\ref{cacat}, namely: 
$I_{\mu+\bar{\mu}}(>1\mbox{ TeV}) <  3.5\  /\ ({\rm km^{2}\ yr})$ at 90 \% CL.
This will make the search for a signal more difficult but will be accompanied
by a decrease of the background, for the sources of high energy neutrinos are the relatively small 
molecular clouds and not the much larger SNR.  

By comparing the size of the overdense 
clouds with the one of the SNR, one would expect in ideal conditions a  
decrease of the background
by a factor of ten; however, the decrease
will be limited by instrumental features,
if the angular  resolution  of the neutrino telescopes will be  larger than the cloud size.
Just for illustration, an angular  resolution of $\delta\theta=0.2^\circ$ at the relevant energies 
corresponds to a search window of 
$\pi\ \delta\theta^2=$40 $\mu$sr. Multiplying by the number of the main 
overdense clouds, i.e.\  four (see Sano {et al.}\ 2010 \cite{sano}) and comparing to 
 the size of the SNR implies a decrease of the background by a factor of two.

We would like to emphasize that the model of Zirakashvili \& Aharonian 2010 \cite{az}
does not necessarily imply a strong reduction of the neutrino
detection rate compared to the pure hadronic model, because in the composite spectrum
the most relevant $\gamma$-rays--those with energy above 10 TeV--are contributed mainly by 
cosmic ray interactions. On the other hand, the composite model implies that the $\gamma$-rays and 
of course the high energy neutrinos are produced in more compact regions, which leads to a significant reduction of background events.

We will be in a better position to quantify the expected, neutrino-induced muon flux
when we will know the results of the analyses of Agile and Fermi. 
In order to predict the very high-energy neutrino flux, 
it would be even more important to know the amount of  
very high-energy $\gamma$-rays correlated with the molecular clouds. 
HESS could provide us with some evidence for 
such a hadronic emission, but future high statistics observations will be crucial 
to obtain reliable measurements (or strong limits) of this component
of the $\gamma$-ray emission.

\section{An appraisal of a telescope in the Northern Hemisphere\label{s1}}
In the last section, we discuss the importance of operating 
a new telescope for high energy neutrinos in the Northern Hemisphere.
The discussion elaborates quantitatively the oft-heard observation: 
most of the Galaxy, being in the Southern Hemisphere, lies in the 
Northern neutrino sky. 
\bigskip

We begin by considering an educated guess on the galactic sources of neutrinos.
It is plausible that the distribution of neutrino sources follows the mass 
distribution of supernovae, young matter and/or star-forming regions. 
We use the distribution of neutron stars of Yusifov \& K\"u\c{c}\"uk 2004 \cite{yk}, 
adopted for the study of supernova neutrinos of Mirizzi {et al.}\ 2006 \cite{mir}
(see their Eqs.~(1), (2) and (4) and compare with Costantini, Ianni, Vissani 2005 \cite{vis}).
We set the $x$-axis from the galactic center to the Earth,
the $z$-axis in the direction of the galactic North, obtaining 
for components of the Earth's rotation axis the unit vector $(u_x,u_y,u_z)=(.484,.747,.456)$.

\begin{figure}
\centerline{\includegraphics[width=0.5\textwidth,angle=0]{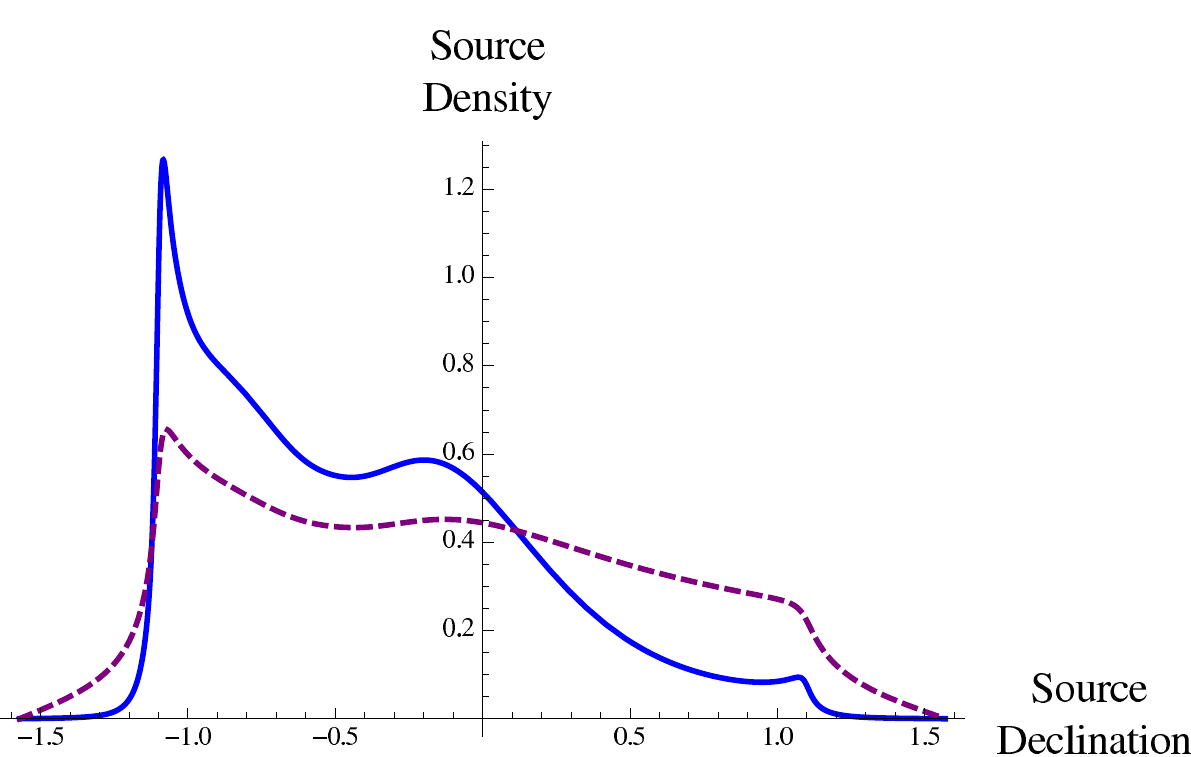}}
\caption{\em Continuous line: 
normalized mass distribution of the Galaxy, as a function of the 
declination. 
Dashed line, the same but weighted with the inverse squared distance from
the mass. 
\label{fig1}}\vskip2mm
\end{figure}

Now we derive 
the normalized mass distribution as a function of the declination of the sources $\delta$, that we
regard as the probability of finding neutrino sources. Similarly, we derive the mass 
distribution weighted with the inverse of the squared distance of the source, 
accounting for the fact 
that the number of events scales  as $1/r^2$ for a standard source.  The results are given in 
Fig.~\ref{fig1} and are easy to understand: The angle between the galactic 
plane and the Celestial equator is $\sim \pi/3$, thus most of the matter is 
at $|\delta|<1$; furthermore,
the galactic center is at $\delta\sim -\pi/6$, thus the region $\delta <0$ is more populated; finally, the 
features are less prominent when we include $1/r^2$ since this emphasizes the local patch of the 
Galaxy rather than the distant regions.

Let us consider the traditional signal of induced muons 
(see Markov 1963 \cite{mrk} for the original references).
High energy neutrino detectors observe 
only downward to safely avoid atmospheric muons;
thus, a source at declination $\delta$ is seen only for a fraction of time:
\begin{equation}
f[\delta,\phi]=1-\frac{{\rm Re}[\arccos(-\tan\delta \tan\phi)]}{\pi}
\label{eff}
\end{equation} 
as a function of the latitude $\phi$ of the detector, 
as discussed {e.g.}, in Costantini \& Vissani 2005 \cite{vis}.
For instance, the galactic center, that is in the Southern sky at 
$\delta=-29^\circ$, is invisible in IceCube ($f=0$) 
and it is seen for a fraction of time $f=$67\%, 63\%, 64\% or 75\% 
in Antares, NEMO, Nestor or Baikal respectively.

By convoluting $f$ with the distribution of the matter in the Milky Way 
we estimate the relevance of a high energy neutrino detector. The  result is shown in the curves 
Fig.~\ref{fig11}. They are symmetric around 1/2 when $\phi\to -\phi$, just as $f$: 
$f[\delta,\phi]+f[\delta,-\phi]=1$, for two antipodal detectors see the entire sky.
From this figure, we verify that the South 
Pole is a less promising place to search for neutrinos from
galactic sources. A detector in the Mediterranean, say with latitude  $\phi=36^\circ 30'$, has 2.9 times better chances; when we weight the mass distribution with $1/r^2$, the improvement is a factor of 1.4
instead. The first number applies if the hypothetical sources are so intense, that all of them can be seen; the second one is plausibly a better estimation of the factor of improvement if there is a sort of ``standard source'' with a fixed intensity, and the neutrino detectors are able to see only the closest ones.

\begin{figure}[t]
\centerline{\includegraphics[width=0.5\textwidth,angle=0]{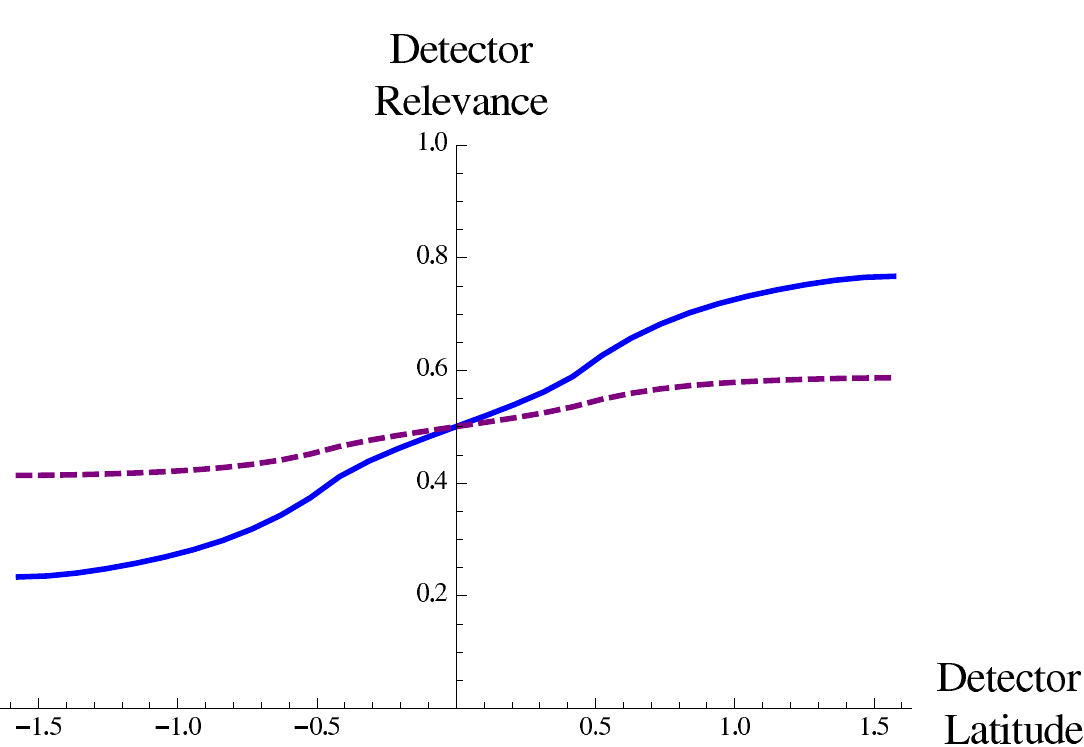}}
\caption{\em Fraction of the Galaxy  seen by a neutrino detector
at  a given latitude. For the dashed lines, the mass distribution was  
weighted with $1/r^2$  to take into account that nearby sources are easier to detect.
\label{fig11}}\vskip2mm
\end{figure}

Similar arguments are frequently invoked in favor of a detector in the 
Northern hemisphere; however, the quantitative evaluation 
of the factor of improvement that we obtained is, to the best of our knowledge, a new result.

\paragraph{Remarks and caveats}
There are other aspects to be kept in mind; {e.g.},  
it seems possible to cover safely a few degrees above the horizon 
already with IceCube (Abbasi {et al.}\ 2009 \cite{ice}). Also,  
when KM3NeT will operate, a portion of the sky will be already 
explored by IceCube; however, some redundancy 
in the observations could be precious 
to cross check the proper functioning of the  detectors.

One can repeat the  calculations including the halo,  or 
a ``bar'' as part of the Milky Way, possibilities that above we 
disregarded.  Alternatively, one could 
consider suitable generalizations, for instance the case of dark matter decay or annihilation; the latter 
requires to weight the fraction $f$ with the square of the density of the dark 
matter distribution. Many of these cases resemble closely a source 
in the galactic center, discussed above.

\section{Summary and discussion}
The discovery of sources of high energy neutrinos is 
a well-recognized scientific  goal:
The hunt has been opened by IceCube and will be complemented 
by  KM3NeT, see Riccobene \& Sapienza 2009 \cite{reov} and 
Anchordoqui \& Montaruli 2009 \cite{rv} for reviews. 
However,  we cannot rely 
on precise predictions yet. 
Clear expectations 
would be helpful or even necessary to focus the search and to optimize the new instruments: 
their area, geometry, energy threshold, etc..
The precise upper bounds that we can derive from 
$\gamma$-ray data are useful, but insufficient for these 
aims. 
With these considerations in mind, 
we made an effort to determine some of the boundaries of the 
present knowledge and to discuss the prospects 
to proceed toward definite expectations, focussing 
on  the main target of KM3NeT: galactic neutrino sources.

A new high-energy neutrino telescope in the Northern Hemisphere is considered 
highly desirable.
We discussed in Sec.~\ref{s2} which $\gamma$-ray sources  
may yield a minimum signal in neutrino telescopes, assuming that  
the $\gamma$-rays are not absorbed.
We found that, for high energy neutrino search, it would be particularly important to know 
the $\gamma$-ray sources with a sufficiently intense emission around 20 TeV: see Eq.~\ref{cat}. 
We argued in Sec.~\ref{s3} that there are reasonable
chances of getting a reliable prediction for  RX J1713.7-3946, that could become a reference target for a telescope located in the Northern hemisphere. The chances are linked to future analyses of 
existing $\gamma$-ray data: those by Agile and Fermi, which should reveal an emission more intense than the one expected assuming the leptonic mechanism; those by HESS above $20-30$ TeV,
that should reveal the correlation of the $\gamma$-ray events 
with the molecular clouds that interact with RX J1713.7-3946. 
Finally, we discussed the importance to have a new high-energy 
neutrino telescope in the Northern Hemisphere. 
We developed in Sec.~\ref{s1} the 
oft-heard argument in favor of such an instrument, 
concluding that it will be superior by a factor of 1.4-2.9 to IceCube
as a monitor of galactic neutrino sources distributed as the matter of the Milky Way.

\subsection*{Acknowledgement}
\noindent{\footnotesize We are grateful to R.~Ruffini for hospitality in ICRANet, Pescara, where
this work was begun. 
We thank P.L.~Ghia, P.~Lipari, A.~Masiero, N.~Paver, G.~Riccobene and O.~Ryashzkaya 
for useful discussions, hints and help. 
F.V.\ thanks M.L.~Costantini and F.L.~Villante for collaboration on related topics.}

\end{twocolumn}

\end{document}